\documentclass[aps,pra,preprint,12pt]{revtex4-1}

\usepackage{amsmath,amssymb}
\usepackage{multirow}
\usepackage{bm}
\usepackage{color}
\usepackage{graphicx}
\usepackage{CJKutf8}

\begin{document}
\begin{CJK*}{UTF8}{gkai}

\title{\large \bf Theoretical study of the spectroscopic properties of mendelevium ($Z=101$)}

\author{Jiguang Li (李冀光)}
\affiliation{Institute of Applied Physics and Computational Mathematics, Beijing 100088, China}

\author{Vladimir Dzuba}
\affiliation{School of Physics, New South Wales University, Sydney 2052, Australia}

\date{\today}

\begin{abstract}
  Using recently developed version of the configuration interaction method for atoms with open shells we calculate electron structure and spectroscopic properties of the mendelevium atom (Md, $Z=101$). These include energy levels, first and second ionisation potentials, electron affinity, hyperfine structure and electric dipole transition amplitudes between ground and low lying states of opposite parity. The accuracy of the calculations is controlled by performing similar calculations for lighter analog of mendelevium, thulium atom and comparing the results with experiment and other calculations. The calculations for Md are to address the lack of experimental data and help in planing and interpreting the measurements.
\end{abstract}

\maketitle
\end{CJK*}

\section{Introduction}

Mendelevium (Md) is a synthetic element produced in a laboratory. It is not found in nature and it has no stable isotopes. Since its discovery in 1955 more than twenty isotopes of Md were synthesised and studied. However, spectroscopic properties of Md are still not known. Recently, ionisation potentials (IP) of four heavy actinides (Fm, Md, No, Lr) were studied both experimentally and theoretically~\cite{Sato2018}. There are also some spectroscopic data for both mendelevium neighbours in the periodic table. Seven transition frequencies of Fm were measured and many more calculated~\cite{Sewtz2003,Backe2005,ErFm}. The strongest electric dipole transition between the $^1$S$_0$ ground state and the $^1$P$_1^{\rm o}$ excited states of No was  measured for three nobelium isotopes ($^{252}$No,$^{253}$No and $^{254}$No)~\cite{No-exp}. The measurements include hyperfine structure (hfs) for $^{253}$No, and isotope shift for all three isotopes. These measurements, when supported by calculations~\cite{No-IS-exp,No-EL-hfs,NoIS}, lead to extraction of nuclear parameters, such as magnetic dipole and electric quadrupole moments of $^{252}$No nucleus and change of nuclear RMS radius between three isotopes. There are plans for similar measurements in Lr~\cite{Lr-plan}. To the best of our knowledge, there are no similar measurements or calculations for Md. The aim of present paper is to address this gap in the data.

We calculate frequencies and transition rates for some electric dipole transitions between ground and excited states of Md, hyperfine structure for the ground state, electron affinities and ionisation potentials. Having theoretical predictions for frequencies and transition rates is important for the planing of the measurements. Comparing theoretical and experimental hyperfine structure would lead to extraction of nuclear moments. In particular, it is important to have the values for nuclear electric quadrupole moments. They are correlated with nuclear deformation parameters. Studying nuclear deformation is relevant to search for stable nuclei since they are expected to be spherically symmetric.
Knowing IP and electron affinities would lit some light on the chemical properties of the element. We perform similar calculations for Tm atom, which is lighter analog of Md and for which a lot of experimental and theoretical data are available. This allows us to control the accuracy of the calculations for Md.

Both atoms, Tm and Md, have complicated electron structure. The ground states are, the $4f^{13}6s^2 \ ^2$F$^{\rm o}_{7/2}$ state for Tm and the $5f^{13}7s^2 \ ^2$F$^{\rm o}_{7/2}$ state for Md. Excited states include excitations from both $f$ and $s$ states. This means that all fifteen external electrons should be treated as valence electrons. This would lead to a matrix eigenvalue problem of unmanageable size in the standard configuration interaction (CI) approach. An alternative approach, which is a version of the CI method in which most of the excited configurations are treated perturbatively (the CIPT method), has been recently developed~\cite{CIPT,FCI} and successfully applied to a number of atoms with open shells (see, e.g. Refs.~\cite{Db18,E118,SHE6d,E110-112,ErFm}). In this work we apply the CIPT method to calculate  the electron structure of Tm and Md. This allows us to make reliable predictions for the spectroscopic properties of Md. We also study the role of correlation and relativistic effects and use it to explain the differences in Tm and Md spectra.

\section{Theory}

In present paper we are dealing with atomic systems which have fifteen external electrons (e.g. $4f^{13}6s^2$ in Tm and $5f^{13}7s^2$ in 
Md). We use the CIPT method~\cite{CIPT} to perform the calculations. The method is a version of the configuration interaction approach.
In the CI method, an atomic state function for valence electrons (ASF) $\Psi$ is the linear combinations of configuration state functions (CSFs) $\Phi$ with the same parity ($P$), total angular momentum ($J$), and its component along the $z$ direction ($M_J$), i.e.,
\begin{equation}
\Psi(P J M_J) = \sum_{i=1}^{NCSF} c_i \Phi_i (P J M_J).
\end{equation}
Here, \textit{NCSF} represents the number of configuration state functions included in the expansion, and $c_i$ is referred to as the mixing coefficient. The latter and corresponding energies $E_i$ can be obtained by solving the eigenvalue equation of the atomic CI Hamiltonian $H^{\rm CI}$,
\begin{equation}
\label{CI-H}
(H^{\rm CI} - E \mathbb{I}) \mathbf{X} = 0,
\end{equation}
where $\mathbb{I}$ is the unit matrix, and the vector $\mathbf{X} = \{c_1, \dotsc, c_{NCSF} \}$. The CI Hamiltonian in atomic unit reads
\begin{equation}
\label{AtomicH}
H^{\rm CI} = \sum_{i=1}^{N} \left[ c {\bm \alpha_i} \cdot {\bm p_i} + (\beta_i - 1) mc^2  + V_{\rm nuc}(r_i) + V_{\rm core}(r_i) \right] + \sum_{n<i}\frac{e^2}{|{\bm r}_n - {\bm r}_i|}. 
\end{equation}
In the equation above, the summation is made over valence electrons, $c$ is the speed of light, and ${\bm \alpha}$, $\beta$ are the Dirac matrices. The two-parameter Fermi distribution is used to generate the nuclear potential $V_{\rm nuc} (r)$. The $V_{\rm core}$ potential is the potential created by electrons in atomic core below the valence subshells.

Each configuration state function is built from Slater determinants which are antisymmetric products of one-electron orbitals. The B-spline technique~\cite{B-spline} is applied to generate these one-electron orbitals in the so-called $V^{N-1}$ approximation, where $V^{N-1}$ is the Dirac-Hartree-Fock (DHF) potential of the ion remained after removing one electron from the valence shell. Specifically, we remove one $6s$ electron in Tm and one $7s$ electron in Md. After self-consistent DHF procedure for the ion is done, the $V^{N-1}$ potential is frozen and used to calculate basis orbitals.
In practice, 40 B-spline states of order nine in a box with radius 40 $a_{\rm B}$ ($a_{\rm B}$ is the Bohr radius) are employed to construct the one-electron orbitals with the angular momentum up to $l=4$.

Configuration state functions are generated by single (S) or double (D) excitations of electrons from the low-energy reference configurations. In case of many-valence-electrons the number of resulting many-electron states and the size of the CI matrix is enormous. In the CIPT approach~\cite{CIPT} the size of the matrix is reduced by many orders of magnitude by neglecting the off-diagonal matrix elements between high energy states. The CI matrix, ordered in terms of the diagonal matrix elements, is divided into four blocks:
\begin{equation}\label{e:CI4}
\mathbf{H^{\rm CI}} = 
\left(
\begin{array}{c|cccccccc}
\\[-0.4cm]
\mathbf{H^{low}}                                &               &                    &   \mathbf{H^{off}}    &   &               \\[0.1cm]
\hline
\multirow{3}{*}{$\mathbf{H^{off}}$}  &  \ddots  &                     &      \mathbf{0}         &   &                \\
                                                          &               &                    &      \ddots                &   &                \\
                                                          &               &  \mathbf{0} &                                 &   &   \ddots  \\
\end{array}
\right) \, .
\end{equation}
The first block is referred to as the low-energy CI matrix $\mathbf{H^{low}}$, since it is associated with the low-energy states which dominate in the ASF. The dimension $N_{\rm low}$ of this block must meet the condition $N_{\rm low} \ll NCSF$, which is required by the CIPT method. All off-diagonal matrix elements between the high-energy states, labeled as $\mathbf{0}$, are neglected. In fact, the contribution of these matrix elements to the energies and wave functions of the low-energy states are indeed insignificant. On the other hand, the diagonal matrix elements for the high-energy states are remained and shown by the dash dots. The blocks $\mathbf{H^{off}}$ denote off-diagonal matrix elements between low- and high-energy states. On the basis of this simplification, the full CI matrix can be reduced to a much smaller effective CI matrix with corrections from $\mathbf{H^{off}}$,
\begin{equation}\label{e:HM}
\mathbf{H^{\rm CI}}  \rightarrow \mathbf{H^{eff}} = \mathbf{H^{low}} + |\mathbf{H^{off}}|^2\mathbf{D},
\end{equation}
where $\mathbf{D}$ is a diagonal matrix containing energy denominators, $d_{ik} = \delta_{ik}(E^{(0)}-E_k)^{-1}$ in which $E^{(0)}$ is an approximation for the energy of the state of interest, $E_k$ is the diagonal matrix element between high-energy states, $E_k =\langle k | H^{\rm CI}|k\rangle$.
Eq.~(\ref{e:HM}) can be rewritten in terms of matrix elements
\begin{equation}
\label{CIPT-matrix-element}
\langle \Phi_i | H^{\rm CI} | \Phi_j \rangle \rightarrow \langle \Phi_i | H^{\rm CI} | \Phi_j \rangle + \sum_{k} \frac{\langle \Phi_i | H^{\rm CI} | \Phi_k \rangle \langle \Phi_k | H^{\rm CI} | \Phi_j \rangle}{E^{(0)} - E_k}.
\end{equation} 
Here, $i,j \leqslant N_{\rm eff} \equiv N_{\rm low}$, $N_{\rm eff} < k \leqslant NSCF$, $E^{(0)}$ converges to $E$ in Eq. (\ref{CI-H}) by iterating calculations of the matrix elements in Eq. (\ref{CIPT-matrix-element}) and diagonalizations of the effective Hamiltonian matrix $\mathbf{H^{eff}}$. Due to the small size of the effective Hamiltonian matrix $\mathbf{H^{eff}}$ ($N_{\rm eff} \ll NCSF$), the CIPT calculation is very efficient, although the summation over high-energy states in Eq. (\ref{CIPT-matrix-element}) costs most of the computer time. 

The Breit interaction is included in the form
\begin{equation}
H^B_{mn} = -\frac{1}{2 r_{mn}} \left[ {\bm \alpha_m} \cdot {\bm \alpha_n} + \frac{({\bm \alpha_m} \cdot {\bm r}_{mn})({\bm \alpha_n} \cdot {\bm r}_{mn})}{r^2_{mn}} \right].
\end{equation}
The quantum electrodynamic (QED) corrections accounting for the Ueling potential and electric and magnetic form factors~\cite{Flambaum2005} are also included. Both, Breit and QED interactions are included on all stages of the calculations, in the self-consistent DHF procedure, in the calculation of B-spline basis states and in the CI Hamiltonian.

To calculate transition amplitudes and hyperfine structure we need to consider interactions of atomic electrons with external fields. There are four types of external fields considered in present work. Interaction of atomic electrons with nuclear magnetic dipole and electric quadrupole moments are needed for calculation of the hyperfine structure,  interaction with laser electric and magnetic fields are needed for calculation of the electric dipole (E1) and magnetic dipole (M1) transition rates. In all cases we use the so called random phase approximation (RPA). The RPA equations have a form (see, e.g. Ref.~\cite{DzuFlaSilSUS87} and references therein)
\begin{equation}\label{e:RPA}
(H^{\rm DHF} -\epsilon_a)\delta \psi_a = - (F + \delta V^{N-1})\psi_a.
\end{equation}
Here, $H^{\rm DHF}$ is the DHF Hamiltonian, index $a$ numerate atomic states in the core, $F$ is the operator of external field, $\delta \psi_a$ is the correction to the core orbital caused by external field, $\delta V^{N-1}$ is the correction to the self-consistent potential caused by the change to all core orbitals.  Equations (\ref{e:RPA}) are solved self-consistently, by iterations,  for all states in the core. Once the convergence is achieved, the renormalised operator $\tilde F = \sum_i (F + \delta V_{\rm core})_i$ is used to calculate matrix elements with the CI wave functions. The summation goes over valence electrons. Hyperfine constants are proportional to the diagonal matrix elements $\langle \Psi_i |\tilde F| \Psi_i \rangle$;  transition amplitudes are given by off-diagonal matrix elements $\langle \Psi_f |\tilde F| \Psi_i \rangle$ (see Refs.~\cite{DzuFlaSUS84,DzuFlaKozPor98,DDF09} for more details).

Electric and magnetic dipole transition probabilities (in s$^{-1}$) from an initial state $i$ to a final state $f$ are given by
\begin{equation}\label{e:TR}
A_{if} = \frac{2.02613 \times 10^{18}}{\lambda^3 (2J_i +1)} |\langle \Psi_f || \tilde F_{\rm D} || \Psi_i \rangle|^2,
\end{equation}
where $\lambda$ (in ${\rm \AA}$) is the wavelengths associated with this transition, $J_i$ the total angular momentum of the initial state, and $\tilde F_{\rm D}$ the electric or magnetic dipole operator (with the RPA correction). 

\section{Results}

\subsection{Low-lying energy structures and associated E1 transition rates in Tm and Md} 

\label{sec:results_energy_E1}

As mentioned earlier, Tm is lighter analog of Md. In order to control the computational uncertainties, we calculated the energy levels of Tm first. In this calculation, the $V^{N-1}$ approximation was applied to generate the one-electron orbitals, that is, the  [Xe]$4f^{13}6s$ configuration of the Tm$^+$  ion was adopted to produce the $V^{N-1}$ potential. The correlations between valence electrons are included by means of the CIPT technique as described above. The many-electron states for the CIPT calculations are generated by exciting one or two electrons from the reference configuration. The [Xe]$4f^{13}6s^2$ ground state configuration was used as a reference configuration to calculate odd parity states. Two reference configurations, [Xe]$4f^{12}5d6s^2$ and [Xe]$4f^{13}6s6p$, were used to calculate even parity states. Only states from the reference configurations are included in the $\mathbf{H^{low}}$ part of the CI matrix (\ref{e:CI4}). All generated states are treated perturbatively; they are included in the last term of (\ref{e:HM}) or 
(\ref{CIPT-matrix-element}).

The resulting energy structures and E1 transition probabilities are presented in Table~\ref{table:EnergyStructure-Tm} and~\ref{table:E1-Tm}. It was found from Table~\ref{table:EnergyStructure-Tm} that the difference in transition energies between our calculation and ones from NIST database~\cite{NIST} are overall around a few percent for the levels concerned, except for $4f^{12}6s^{2}5d$, $4f^{12}6s^{2}5d$, and $4f^{12}6s^{2}5d$ levels where the deviation reaches about 10\%. Land\'e g-factors are also displayed in Table~\ref{table:EnergyStructure-Tm} providing useful information for states identification. The only a noticeable  discrepancy was found for states 12 and 13 of the $4f^{12}6s^{2}5d$ configuration.

The electric dipole transition probabilities, the corresponding wavelengths and the line strengths are listed in Table~\ref{table:E1-Tm} as well as other measurements available~\cite{Wickliffe1997, Penkin1976}. 

Using similar computational strategy to the case of Tm, we calculated the low-lying energy structures and the E1 transition probabilities for Md ($Z=101$) and presented these results in Table~\ref{table:EnergyStructure-Md} and~\ref{table:E1-Md}. To the best of our knowledge, there are no other theoretical or experimental data available for comparison for this atomic system. However, the uncertainties should be consistent with its homologous Tm~\cite{Fricke1993, Yu2008, Chang2010}.

It is worth noting the predominant difference in the low-lying energy structures between the Md and Tm atoms. As can be seen from Table~\ref{table:EnergyStructure-Md}, most of levels belonging to the $5f^{13}7s7p$ configuration are lower than those in the $5f^{13}6d7s^2$ configuration. The reason is the stronger relativistic effects in the superheavy elements, which cause radial contraction and energy stabilization for the $j=1/2$ orbitals and have opposite effect for the $j>3/2$ orbitals in outer shells~\cite{Fricke1984, Pyykko1988}. In addition, the fine-structure splitting in the ground configuration of Md I is two times larges than that in Tm I.

\begin{table}[!ht]
\caption{\label{table:EnergyStructure-Tm} Transition energies (in cm$^{-1}$) and Land\'e g-factors of states in the ground and low-lying excited configurations for Tm I.}
\begin{tabular}{cccccccccccccccc}
\hline 
\hline
          &                                     &                                &        \multicolumn{3}{c}{$\Delta E$}        &&     \multicolumn{2}{c}{$g$}            \\
\cline{4-6}\cline{8-9}No.   &    Configuration      &   Term    &  This work  &  NIST~\cite{NIST}  &    Diff.(\%)  &&   This work  &  NIST~\cite{NIST}    \\
\hline
1       &    $4f^{13}6s^{2}$         &    $^2\textrm{F}^o_{7/2}$         &     0             &      0         &     0                     &&     1.1429      &   1.14119     \\
2       &    $4f^{13}6s^{2}$         &    $^2\textrm{F}^o_{5/2}$         &     8382       &    8771     &    389~(4\%)       &&     0.8571      &   0.855         \\
3       &    $4f^{12}6s^{2}5d$    &    $^6\textrm{G}_{9/2}$             &     12780     &   13120    &    340~(3\%)        &&    1.3027       &   1.305         \\
4       &    $4f^{12}6s^{2}5d$    &    $^2\textrm{F}_{7/2}$              &     15574     &   16957    &   1383~(8\%)       &&    1.1612       &    1.1722      \\
5       &    $4f^{12}6s^{2}5d$    &    $^2\textrm{G}_{9/2}$             &     16640     &   18837     &   2197~(12\%)    &&    1.1277      &   1.1318       \\
6       &    $4f^{13}6s6p$          &    $^{12}\textrm{I}_{7/2}$           &     16927     &   16742    &   -185~(1\%)       &&    1.3120       &    1.325       \\
7       &    $4f^{13}6s6p$          &    $^{8}\textrm{H}_{7/2}$           &     17453     &   17343     &  -110~(1\%)       &&    1.0235       &   1.02153     \\
8       &    $4f^{13}6s6p$          &    $^{4}\textrm{G}_{9/2}$           &     17754    &    17614     &  -140~(1\%)       &&   1.1877        &    1.18598    \\
9       &    $4f^{13}6s6p$          &    $^{2}\textrm{D}_{5/2}$           &      17867    &   17753     &  -114~(1\%)       &&    1.1807       &   1.186         \\
10     &    $4f^{13}6s6p$          &    $^{8}\textrm{I}_{3/2}$            &      19110    &    19132     &   22~($<$1\%)   &&    0.8699       &    0.88          \\
11      &    $4f^{13}6s6p$         &    $^{4}\textrm{F}_{5/2}$           &      19485    &    19549     &   64~($<$1\%)   &&    1.0144       &      0.983      \\
12     &    $4f^{12}6s^{2}5d$    &    $^{12}\textrm{I}_{5/2}$          &      20061    &     20407    &   346~(2\%)       &&    1.1277       &     1.58         \\
13     &    $4f^{12}6s^{2}5d$    &    $^8\textrm{G}_{3/2}$             &      21871    &     21799    &   -72~($<$1\%) &&    0.9637       &    1.45          \\
14     &    $4f^{12}6s^{2}5d$    &    $^2\textrm{D}_{3/2}$             &      26169    &     23574    &   -2595~(11\%)  &&    0.8468       &       0.83       \\
\hline        
\end{tabular}
\end{table}

\begin{table}[!ht]
\caption{\label{table:E1-Tm} Wavelength ($\lambda$ in nm), line strength ($S$ in a.u.) and probabilities ($A$ in s$^{-1}$)of E1 transitions between the excited states ($e$) and the states in the ground configuration ($g^o$) for Tm I. The numbers in the first column correspond to that in Table \ref{table:EnergyStructure-Tm}. The numbers in square brackets represent the power of 10.}
\begin{tabular}{lccccccccccccccc}
\hline 
\hline
 &  \multicolumn{2}{c}{$\lambda$}    &&              &&  \multicolumn{2}{c}{$A$}       \\
\cline{2-3}\cline{7-8} $g^o-e$  &  This work   &  Others     &&   $S$    &&      This work   &  Others       \\
\hline
$1-9$     &    559.69     &   563.141$^a$    &&    1.2486        &&  2.40[6]  &  1.09[6]$^a$    \\
              &                    &   563.297$^b$    &&                       &&               &  1.17[6]$^b$    \\
$1-11$   &    513.22     &   511.397$^a$    &&    1.4983[-1]   &&  3.74[5]  &  2.41[5]$^a$    \\
              &                    &   511.539$^b$    &&                        &&              &  3.09[5]$^b$    \\
$1-12$   &    498.48     &   490.032$^b$    &&    6.3382[-2]   &&  1.73[5]  &                        \\
$1-4$     &    642.10     &   589.563$^a$    &&    8.0594[-1]   &&  7.71[5]  &   6.51[5]$^a$  \\
              &                    &   589.727$^b$    &&                        &&              &                        \\
$1-6$     &    590.77     &   597.292$^b$    &&    2.3310[-1]   &&  2.86[5]  &                        \\
$1-7$     &    572.97     &   576.429$^a$    &&    1.1366        &&  1.53[6]  &   3.88[5]$^a$   \\
              &                    &   576.589$^b$    &&                       &&               &  4.21[5]$^b$    \\
$1-3$     &    782.47     &   762.218$^b$    &&    8.8697[-4]   &&  3.75[2]  &                        \\
$1-5$     &    600.96     &   530.712$^a$    &&    2.2611        &&  2.11[6]  &   2.17[6]$^a$   \\
              &                    &   530.859$^b$    &&                       &&               &   2.27[6]$^b$   \\
$1-8$     &    563.25     &   567.584$^a$    &&    1.8012        &&  2.04[6]  &   1.30[6]$^a$   \\
              &                    &   567.741$^b$    &&                       &&               &  1.31[6]$^b$    \\
$2-10$   &    932.14     &   965.158$^b$    &&    1.2202[-3]   &&  7.63[2]  &                        \\
$2-13$   &    741.34     &   767.569$^b$    &&    6.4272[-5]   &&  7.99[1]  &                        \\
$2-14$   &    562.21     &   675.531$^b$    &&    8.5701[-1]   &&  2.44[6]  &                        \\
$2-9$     &   1054.30    &   1113.413$^b$  &&    7.8882[-4]   &&  2.27[2]  &                        \\
$2-11$   &    900.66     &   927.851$^b$    &&    2.9943[-4]   &&  1.38[2]  &                        \\
$2-12$   &    856.24     &   859.432$^b$    &&    6.0700[-3]   &&  3.27[3]  &                        \\
$2-4$     &   1390.43    &   1221.633$^b$  &&    1.0648[-3]   &&  1.00[2]  &                        \\
$2-6$     &   1170.28    &   1254.549$^b$  &&    5.6017[-4]   &&  8.85[1]  &                        \\
$2-7$     &   1102.41    &   1166.571$^b$  &&    4.7888[-3]   &&  9.05[2]  &                        \\
\hline        
\end{tabular} \\[0.1cm]
$^a$ Ref.~\cite{Wickliffe1997} \\
$^b$ Ref~\cite{Penkin1976}
\end{table}

\begin{table}[!ht]
\caption{\label{table:EnergyStructure-Md} Transition energies (in cm$^{-1}$) and Land\'e g-factors of states in the ground and low-lying excited configurations for Md I.}
\begin{tabular}{cccccccccccccccc}
\hline 
\hline
&&&\multicolumn{2}{c}{This work}   \\
\cline{4-5}No.   &    Configuration      &   Term           &        $\Delta E$     &    $g$           \\
\hline
1       &    $5f^{13}7s^{2}$        &    $^2\textrm{F}^o_{7/2}$          &        0                    &     1.1429      \\
2      &    $5f^{13}7s^{2}$         &    $^2\textrm{F}^o_{5/2}$          &        15712             &     0.8572     \\
3       &    $5f^{13}7s7p$          &    $^4\textrm{F}_{7/2}$              &        19223            &    1.2481      \\
4       &    $5f^{13}7s7p$          &    $^8\textrm{H}_{7/2}$             &        20174             &    1.0750     \\
5       &    $5f^{13}7s7p$          &    $^8\textrm{H}_{9/2}$             &        20647            &    1.1943      \\
6       &    $5f^{13}7s7p$          &    $^{12}\textrm{I}_{5/2}$           &        21013             &    1.1151       \\
7       &    $5f^{13}7s7p$          &    $^2\textrm{D}_{3/2}$             &        25149             &    0.8659    \\
8       &    $5f^{13}7s7p$          &    $^{14}\textrm{J}_{5/2}$          &        25548           &    1.0736      \\
9       &    $5f^{13}7s7p$          &    $^6\textrm{G}_{9/2}$             &        25613            &    1.2697      \\
10     &    $5f^{13}7s7p$          &    $^4\textrm{F}_{7/2}$              &        25764            &    1.2031      \\
11      &    $5f^{13}7s7p$         &    $^2\textrm{G}_{9/2}$             &        28796            &    1.1288       \\
12     &    $5f^{13}7s7p$          &    $^{12}\textrm{I}_{5/2}$           &        28817            &    1.1625       \\
13     &    $5f^{13}7s7p$          &    $^{12}\textrm{I}_{3/2}$           &        36534           &    0.7038     \\
14     &    $5f^{12}6d7s^2$      &    $^6\textrm{F}_{3/2}$              &        37857           &     1.0541      \\
\hline        
\end{tabular}
\end{table}

\begin{table}[!ht]
\caption{\label{table:E1-Md} Wavelength ($\lambda$ in nm), line strength ($S$ in a.u.) and probabilities ($A$ in s$^{-1}$) of E1 transitions between the excited states (e) and the states in the ground configuration (g) for Md I. The numbers in the first column correspond to that in Table \ref{table:EnergyStructure-Md}. The numbers in square brackets represent the power of 10.}
\begin{tabular}{lccccccccccccccc}
\hline 
\hline
$g^o~-~e$  &    $\lambda$    &     $S$           &    $A$     \\
\hline
$1~-~6$     &       475.90       &   6.0079         &  1.88[7]  \\
$1~-~8$     &       391.42       &   5.2940[-1]    &  2.98[6]  \\
$1~-~12$   &       347.02       &   4.4985[1]     &  3.64[8]  \\
$1~-~3$     &       520.21       &   4.6559[-1]    &  8.38[5]  \\
$1~-~4$     &       495.69       &   6.5546         &  1.36[7]  \\
$1~-~10$   &       388.14       &   7.9123[-1]    &  3.43[6]  \\
$1~-~5$     &       484.33       &   1.0604[1]     &  1.89[7]  \\
$1~-~9$     &       390.44       &   8.1812[-3]    &  2.78[4]  \\
$1~-~11$   &       347.27       &   6.6227[1]     &  3.20[8]  \\
$2~-~7$     &       1059.66     &   3.1645[-3]    &  1.35[3]  \\
$2~-~13$   &       480.26       &   4.1412         &  1.89[7]  \\
$2~-~14$   &       451.57       &   1.5198[-2]    &  8.36[4]  \\
$2~-~6$     &       1886.44     &   5.3013[-3]    &  2.67[2]  \\
$2~-~8$     &       1016.67     &   5.7567[-3]    &  1.85[3]  \\
$2~-~12$   &       763.07       &   9.7281[-3]    &  7.39[3]  \\
$2~-~3$     &       2848.19     &   1.4457[-3]    &  1.58[1]  \\
$2~-~4$     &       2241.15     &   4.4171[-3]    &  9.94[1]  \\
$2~-~10$   &       994.83       &   4.9412[-5]    &  1.27[1]  \\

\hline        
\end{tabular}
\end{table}

The difference in the spectra of Tm and Md can be understood by looking at the energies of lowest one-electron basis states. These energies, calculated in the DHF approximation, are presented in Table~\ref{t:hfen}. One can see that while moving from lighter atom Tm to heavier atom Md, the energy of the lowest $s$ state goes down by about 0.02~a.u., energies of the lowest $f$ states go up by about the same value while the change in energies of other states is significantly smaller. This is the manifestation of the relativistic effects, which  act directly on the $s$ states and indirectly, through the exchange with the $s$ electrons in the core, on $f$ states. This translates into differences in the spectra of Tm and Md. For example, the $5f^{13}7s7p$ configuration has more $f$ electrons and less $s$ electrons than the $5f^{12}7s^26d$ configuration. Therefore, corresponding states of Md are higher on the energy scale than similar states of Tm. 

\begin{table}[!ht]
\caption{\label{t:hfen}Energies (in a.u.) of lowest one-electron basis states of Tm and Md obtained in the DHF approximation.}
\begin{tabular}{llcll}
\hline
\hline
\multicolumn{2}{c}{Tm}&&
\multicolumn{2}{c}{Md}\\
\cline{1-2}\cline{4-5}\multicolumn{1}{c}{State}&
\multicolumn{1}{c}{Energy}&&
\multicolumn{1}{c}{State}&
\multicolumn{1}{c}{Energy}\\
\hline
$6s$           & -0.411 && $7s$          & -0.431 \\
$6p_{1/2}$ & -0.122 && $7p_{1/2}$ & -0.125 \\
$6p_{3/2}$ & -0.114 && $7p_{3/2}$ & -0.106 \\
$5d_{3/2}$ & -0.082 && $6d_{3/2}$ & -0.080 \\
$5d_{5/2}$ & -0.081 && $6d_{5/2}$ & -0.079 \\
$4f_{5/2}$  & -0.770 && $5f_{5/2}$  & -0.785 \\
$4f_{7/2}$  & -0.719 && $5f_{7/2}$  & -0.698 \\
\hline        
\end{tabular}
\end{table}

\subsection{Hyperfine structure of Tm and Md}

We have calculated hyperfine structure of the ground states of Tm and Md. The results are presented in Table~\ref{t:hfs}.
Thulium atom has only one stable isotope $^{169}$Tm. Its nuclear spin is $I=1/2$ and nuclear magnetic dipole moment $\mu = -0.2316 \mu_N$~\cite{NIST}. This means that it has no electric quadrupole moment contribution to the hfs. However, to test the calculations, having only magnetic dipole contribution is sufficient since the same wave function is used for calculating both magnetic dipole and electric quadrupole contributions. Experimental hfs for $^{169}$Tm is presented as hfs splitting~\cite{Kolachevsky2007}. The splitting, $\Delta E$ is related to the magnetic dipole hfs constant $A$ by $\Delta E = AF_{\rm max}$, where $F_{\rm max} = I+J=4$. We have excellent agreement between theory and experiment (see Table~\ref{t:hfs}).
This gives us confidence that the results for Md are also likely to be accurate. More than twenty isotopes of Md are known, many of them have nuclear spin $I>1/2$ which is necessary for having electric quadrupole moment. Therefore, we have calculated both, magnetic dipole and electric quadrupole hfs constants $A$ and $B$ for Md. The results are presented in Table~\ref{t:hfs}.

\begin{table}[!ht]
\caption{\label{t:hfs}Hyperfine structure (in MHz) of the ground states of $^{169}$Tm and Md isotopes. Magnetic dipole moment of the  $^{169}$Tm nucleus is $\mu = -0.2316 \mu_N$, electric quadrupole moment $Q$ is in barn (1barn=$10^{-28} \ {\rm m}^2$).}
\begin{tabular}{cccc cccc}
\hline
\hline
\multicolumn{1}{c}{Atom}&
\multicolumn{1}{c}{$I$}&
\multicolumn{1}{c}{$J$}&
\multicolumn{1}{c}{$F_{\rm max}$}&
\multicolumn{1}{c}{$A$}&
\multicolumn{1}{c}{$B$}&
\multicolumn{1}{c}{$\Delta E_{\rm theor}$}&
\multicolumn{1}{c}{$\Delta E_{\rm expt}$~\cite{Kolachevsky2007}}\\
\hline
Tm & 1/2 & 7/2 & 4 & -359 & 0 & -1436 & -1496.550(1) \\
Md &       & 7/2 &    & 826($\mu/I$) & -1808$Q$ &  &  \\
\hline        
\end{tabular}
\end{table}

\subsection{Lifetime of the $4f^{13}6s^{2}~^2\textrm{F}^o_{5/2}$ state}

Since the M1 transition between the fine-structure levels in the $4f^{13}6s^{2}$ ground state configuration of Tm is considered as a candidate for a clock transition~\cite{Kolachevsky2007}, the lifetime of the $4f^{13}6s^{2}~^2\textrm{F}^o_{5/2}$ state was calculated in~\cite{Kolachevsky2007} in two different approximations. We calculate it too, starting from solving the RPA equations (\ref{e:RPA}) and using then the formula (\ref{e:TR}) for transition rates.
Our result is displayed in Table~\ref{Table:2F_Lifetime} together with previous calculations performed with the use of the Cowan and FAC codes. It can be seen that all theoretical values are consistent with each other~\cite{Kolachevsky2007}. 

\begin{table}[!ht]
\caption{\label{Table:2F_Lifetime} Lifetime (in s) of the $4f^{13}6s^{2}~^2\textrm{F}^o_{5/2}$ state in Tm I.}
\begin{tabular}{ccccc}
\hline
\hline
Level                                                        &     This work         &      Others                       \\
\hline
$4f^{13}6s^{2}~^2\textrm{F}^o_{5/2}$     &       0.147             &      0.17$^a$, 0.13$^b$   \\ 
\hline
\end{tabular}\\[0.1cm]
$^a$ Calculated by Cowan code~\cite{Kolachevsky2007} \\
$^b$ Calcualted by FAC code~\cite{Kolachevsky2007} \\
\end{table}

\clearpage
\subsection{Ionization potential}
Although the ``$s$" orbitals in the outermost shell shrink due to the relativistic effects, as discussed earlier in Section~\ref{sec:results_energy_E1}, the ground state is still [Xe]$4f^{13}6s \ (7/2,1/2)^{\rm o}_4$ for Tm II and [Rn]$5f^{13}7s \ (7/2,1/2)^{\rm o}_4$ for Md II, i.e., ionization goes with removal of $s$ electron. This is apparent from the one-electron energies calculated in the DHF approximation (see Table~\ref{t:hfen}). The $f$ states are significantly lower on the energy scale than $s$ states in both atoms.  IP is calculated as a difference between CIPT energies of the ground states of the neutral atom and its ion. We use the same $V^{N-1}$ potential in both cases, so that basis orbitals remain the same. This ensures exact cancellation between energies associated with remaining electrons. The calculated ionization potentials are presented in Table~\ref{IP}. There is a lot of theoretical works on the ionization potential of the lanthanide and actinide atoms~\cite{Zheng1991, Tatewaki1995, Sekiya1997, Liu1998, Sekiya2000, Cao2003, Gaigalas2004}. Reviewing this works would go beyond the scope of the present paper. Since we need to estimate the uncertainties of our calculations, we compare our results with experiment only~\cite{Zmbov1966, Hertel1968, Martin1974,  Ackermann1976,  Worden1978, Wendt2014a, Sato2018}. Comparing with the most accurate experimental values so far, IP~(Tm I) = 6.18436(6)~eV obtained from spectroscopic data~\cite{Martin1974} and IP~(Md I) = 6.59(13)~eV measured by a surface ionization process coupled to an online mass separation technique in an atom-at-a-time regime~\cite{Sato2018}, we estimated our computational uncertainties at about 1\%. 

In addition, the fine-structure splittings in the ground state of Tm II and Md II are also displayed in Table~\ref{IP}. Good agreement was found for the excitation energies for Tm II between ours and those from NIST database~\cite{NIST}.

The ionization potentials for the singly charged Tm$^{+}$ and Md$^{+}$ were also calculated (see Table~\ref{IP_II}). 
In this calculations the $V^{N-2}$ potential, corresponding to removal of one more $s$-electron, was used to generate the basis orbitals. The difference in IP for Tm II is $\sim0.5$~eV (or $\sim$ 5\%) between our calculation and other results~\cite{Martin1974, NIST}. This uncertainty is also expected for the case of Md$^{+}$. The M1 transition energies in the $nf^{13}$ ground configuration of Tm III and Md III, where $n=4$ for the former and $n=5$ for the latter, are reported in Table~\ref{IP_II}. For Tm III our result is consistent with the one from NIST database~\cite{NIST}.

\begin{table}[!ht]
\caption{\label{IP} Ionization potentials (IP in eV) of the neutral atoms Tm and Md and transition energies ($\Delta E$ in cm$^{-1}$) in the ground state $4f^{13}6s$ for Tm II and $5f^{13}7s$ for Md II.}
\begin{tabular}{cccccccccccccccc}
\hline 
\hline
                    &                          &&           \multicolumn{3}{c}{ $\Delta E$}          											     \\
\cline{4-6}     &   IP($^3\!F_4$)  &&    $^1\!F_3$    &      $^3\!F_2$     &     $^3\!F_3$  								          \\
\hline
\multicolumn{6}{c}{Tm II}  \\
This work                                                           &       6.24                 &&       243          &       8321         &       8515           \\
Zmbov and Margrave~\cite{Zmbov1966}             &   5.87 $\pm$ 0.01   &                               							      \\
Hertel~\cite{Hertel1968}                                     &   6.03 $\pm$ 0.04                                                        				        \\
Martin \textit{et al.}~\cite{Martin1974}                 &   6.18436(6)            &                                    						      \\
Ackermann \textit{et al.}~\cite{Ackermann1976}  &   6.11 $\pm$ 0.1     &                     									       \\
NIST                                                                 &       6.18                 &&       237          &       8770        &       8957            \\
\hline
\multicolumn{6}{c}{Md II}																								      \\
This work  										 &       6.67                 &&       381          &       15629       &       15943          \\
Martin \textit{et al.}~\cite{Martin1974}  		   &       6.58(7)             &                                      						        \\
Sato \textit{et al.}~\cite{Sato2018}                      &   6.59 $\pm$ 0.13   &                      								         \\
NIST       									         &       6.58                 &&                       &                        &                        \\
\hline        
\end{tabular}
\end{table}

\begin{table}[!ht]
\caption{\label{IP_II} Ionization potentials (IP in eV) of Tm$^{+}$ and Md$^{+}$ as well as fine-structure splittings ($\Delta E$ in cm$^{-1}$) in the corresponding ground state $4f^{13}$ for Tm$^{2+}$ and $5f^{13}$ for Md$^{2+}$.}
\begin{tabular}{cccccccccccccccc}
\hline 
\hline
                       &   IP($^2F_{7/2}$)  &    $\Delta E$ ($^2F_{7/2}~-~^2F_{7/2}$2)      \\
\hline
\multicolumn{3}{c}{Tm III}                                                                                        \\
This work(I)    &          11.54            &                    8332                                          \\
Martin \textit{et al.}~\cite{Martin1974}   &   12.05(8)   &                                            \\
NIST              &          12.065          &                    8774                                          \\
\hline
\multicolumn{3}{c}{Md III}                                                                                       \\
This work(I)    &          12.30            &                   15682                                        \\
\hline        
\end{tabular}
\end{table}

\clearpage

\subsection{Electron affinity}

An initial hypothesis of the ground state would be $4f^{14}6s^2$ for Tm$^-$ and $5f^{14}7s^2$ for Md$^-$. However, the energy 
of the $f$ orbital is not low enough to be bound due to the relativistic effect. In 1994 Chevary and Vosko predicted, based on 
a Dirac-Hartree-Fock functional density theory study of the 70 electron system, that the ground state of Tm$^-$ should be $4f^{13}6s^26p~^5\!F_3$~\cite{Chevary1994}. Moreover, they estimated the electron affinity to be in the range from 1 to 5
 mHartree~\cite{Chevary1994}. Later, this theoretical prediction was confirmed by an experiment using the electric dissociation
  technique, and the electron affinity, EA=0.032(7)~eV, was measured for Tm$^-$~\cite{Nadeau1994, Nadeau1997}. Recently,
 Davis and Thompson reported an experimental value with smaller uncertainties, EA=1.029 $\pm$ 0.022~eV~\cite{Davis2001}. 
 This result was obtained using laser photodetachment electron spectroscopy for Tm$^-$. Its value is almost two order 
 of magnitude larger than others. Later, O'Malley and Beck reported a new result of calculations, EA=0.022~eV~\cite{Tm-} which is consistent with the first predictions and measurements.
 
We use the CIPT method to evaluate the electron affinity of Tm I. Calculations are similar to calculation of IP. EA is found as a difference between ground state energies of the neutral atom and its negative ion. The same $V^{N-1}$ potential is used in both cases. The results are presented in Table~\ref{EA}. As one can see, our electron affinity is in better agreement with those by Chevary and Vosko~\cite{Chevary1994}, Nadeau \textit{et al.}~\cite{Nadeau1994, Nadeau1997} and O'Malley and Beck~\cite{Tm-}. In addition, our result for the fine-structure splitting between two bound states of Tm$^-$ agrees well with Chevary and Vosko result in the range from 5.44~meV to 8.16~meV~\cite{Chevary1994}, and experimental values by Nadeau \textit{et al.}, $\Delta E = 5$~meV~\cite{Nadeau1994} and 7~meV~\cite{Nadeau1997}, but still differs from the latest measurement of 50~meV~\cite{Davis2001}.

For the case of Md, the $5f^{13}7s^27p~^5\!\textrm{F}_3$ ground state of its negative ion was obtained based on the CIPT calculation, and the electron affinity was predicted to be 0.169~eV. Similar to Tm$^-$, there is another bound state $5f^{13}7s^27p~^3\!\textrm{G}_4$ with an energy interval around 10~meV to the ground state of Md$^-$.

\begin{table}[!ht]
\caption{\label{EA} Electron affinity (EA in eV) for Tm$^{-}$ and Md$^{-}$.}
\begin{tabular}{cccccccccccccccc}
\hline 
\hline
                        &                                   &                                   &              \multicolumn{3}{c}{EA}   \\         
\cline{4-6} Ion   &        Configuration        &       Term                &     This work  &  Other theory & Experiment \\
\hline
Tm$^{-}$         &   $4f^{13}6s^{2}6p$    &   $^5\textrm{F}_3$     &        0.094        &   0.027$\thicksim$0.136~\cite{Chevary1994} &
                                                                                                                       0.032$\pm$ 0.007~\cite{Nadeau1994, Nadeau1997}  \\
                       &                                    &                                   &      &0.022~\cite{Tm-}   &   1.029$\pm$ 0.022~\cite{Davis2001} \\
                       &                                    &   $^3\textrm{G}_4$     &        0.081        &     & \\
Md$^{-}$         &   $5f^{13}7s^{2}7p$    &   $^5\textrm{F}_3$      &        0.169       &      & \\
                       &                                    &   $^3\textrm{G}_4$     &        0.159       &      & \\
\hline 
\end{tabular}
\end{table}

\section{Conclusion}

In this work, we employed the CIPT method to calculate the energy structures and E1 transition probabilities between the low-lying excited states and states in the ground configuration for the superheavy element Md I as well as its homologous Tm I. In addition, the first and second ionization potentials, hyperfine structure and the electron affinities were also reported for these two elements. Good agreement were found between our results and others when available for all atomic properties under investigation, with an exception of the electron affinity for Tm I. The present value differs from the latest measurement by a factor of 10, although it is consistent with other calculations and previous measurements. The two bound states of Md$^-$ were found by the calculations, and the electron affinity was estimated to be EA=0.169~meV for Md I.

\acknowledgements

This work was supported by the National Natural Science Foundation of China (Grant No. 11874090) and the Australian Research Council. V.A.D. would like to express special thanks to the Institute of Applied Physics and Computational Mathematics in Beijing 
for its hospitality and support.

\bibliography{SHE_other,SHE_dzuba,SHE}

\end{document}